\documentclass[12pt]{article}
\usepackage{amsmath}

\begin{document}

\title{Partially reduced formulation of scalar Yukawa model:
Poincar\'e-invariance and unitarity}
\author{I Zahladko$^1$ ~and~ A Duviryak$^2$\\
Institute for Condensed Matter Physics of NAS of Ukraine,\\
Lviv, UA-79011, Ukraine \\$^1$zagladko@icmp.lviv.ua,
$^2$duviryak@icmp.lviv.ua}

\maketitle

\begin{abstract}
We consider a scalar Yukawa-like model in the framework of partially
reduced quantum field theory. The reduced Lagrangian of the model
consists of free scalar field terms and nonlocal current interaction
term. Hamiltonian expressions for conserved quantities arose from a
Lorentz-invariance of the model in the momentum representation have
been found in the first-order  approximation with respect to a
coupling constant squared. Canonical quantization of the system is
performed. It is shown that the obtained conserved quantities and
previously founded the Hamiltonian and the momentum of the system
satisfy the commutational relations of the Poincar\'e group. The
expression for S-matrix in the current approximation is found.
Unitarity of this operator is proven by direct calculation.
\end{abstract}


\maketitle


\section {INTRODUCTION}

Recently a partially reduced field theory \cite{Dar98}--\cite{du}
complemented with the variational method \cite{Ste85,dar} are being
used for a description of the relativistic bound states problem
\cite{D-D02}--\cite{D-D10}. The structure of this approach is as
follows. Variables of a field mediating an interaction of fermion or
scalar matter fields are eliminated from the Lagrangian of the
system by means of covariant Green function, the propagator of
mediating field. The reduced Lagrangian description is put into the
Hamiltonian form which then is quantized canonically. Finally, a
field-theoretical version of a variational method is applied in
order to derive relativistic wave equations describing bound or/and
scattering states of a system.

    A reduced Lagrangian includes space-time-nonlocal interaction term.
Unlike other nonlocal theories known in literature where a
nonlocality is inserted by hands (in a free-field Lagrangian
\cite{P-U50,Ef} or in interaction terms by means of form-factors
\cite{Bl_p,Ef}), here the nonlocality appears in natural way, via a
propagator mediating an interaction between currents of matter
field. Thus the reduced field theory describes those processes of an
original local theory in which a role of free quanta of mediating
field can be neglected. This approach has been used to the
description of positronium (Ps), muonium (Mu) \cite{D-D02,T-D04}
Ps$^-$ and Mu$^-$ \cite{B-D08}, and obtained spectra agree with
conventional QED and experimental data. The reduced scalar Yukawa
model \cite{Dar98,Dar00,du,ERD05} and its nonlinear generalizations
\cite{D-D10} were considered too. In all cases it took sparing
efforts to derive the variational wave equations of advantageous
structure as to compare to the Bethe-Salpeter (BS) equations.

    However, some important problems of the partially reduced field theory
remain unconsidered. A nonlocality of the Lagrangian complicates a
transition to the Hamiltonian formalism. We apply a Hamiltonization
scheme developed by Llosa and Vives \cite{los} for nonlocal
Lagrangians in mechanics. This procedure is realized by the
subsequent approximation scheme and leads to a loss of covariance
\cite{du}. Non-covariance and non-exactness of this method cause to
distrust in a relativistic invariance of the approach and thus in
its physical meaningfulness. It is known that covariance is not a
necessary condition of Poincar\'e-invariance of the system, but a
Poincar\'e-invariance itself has a physical sense and it is a
necessary condition of a reliability of results predicted by the
theory.

In present paper this problem is considered for the simple scalar
Yukawa model. We construct the Hamiltonian formulation of a reduced
Yukawa-like model in the linear (i.e., second-order coupling
constant) approximation and prove a Poincar\'e-invariance of the
model. Ten generators of the Poincar\'e group have been constructed
for this purpose. They are built on the basis of Noether currents,
by means of transition to the Hamiltonian formalism (Sections 2-4)
and further quantization (Section 5). It is worth mentioning that
not only Hamiltonian but also boost generator contains the
interaction term. It is shown that these generators satisfy
commutational relations of the Poincar\'e group within the limits of
present approximation (Section 6).

The other important problem inherent to nonlocal field theories
is the construction of unitary scattering matrix.
Usually, difficulties that arise herewith are the
reasons to distrust such theories. In the Subsection 7.1 we
construct the scattering matrix of the reduced Yukawa-like model
by means of the standard quantum-mechanical algorithm \cite{Qm},
using a transition to the interaction representation.
Unitarity of scattering matrix within the present approximation
is shown in Subsection 7.2. Some details of this computation are
given there too.

We use the time-like Minkowski metrics: $\|\eta_{\mu\nu}\|={\rm
diag}(+,-,-,-)$, and put $c=\hbar=1$.


\section {REDUCED LAGRANGIAN AND CONSERVED QUANTITIES}

The considered model comes from the scalar Yukawa model
\cite{Dar98}, which describes the dynamics of two complex scalar
fields $\phi_r (x)$, $(r=1,2)$ coupled via a real scalar mediating
field $\chi(x)$.

Reduction of the field $\chi(x)$ in the initial Lagrangian of the
Yukawa model leads to an effective non-local Lagrangian describing
the interaction of currents of fields $\phi_r (x)$ in terms of the
symmetric Green function of Klein-Gordon equation \cite{Dar98,du}.
For generality we replace the Green function by an arbitrary
symmetric Poincar\'e-invariant kernel, $K(x-x')=K(x'-x)$.

Hence a starting point of our work is a Lagrangain density:
%
\begin{equation}
\label{L}
L=\sum\limits_{r=1}^2 L_r +\frac{1}{2}\int d^4x \rho(x)K(x-x')\rho(x'),
\end{equation}
where
%
\begin{equation}
L_r=(\partial_\mu \phi_r^*)(\partial^\mu \phi_r)-m_r^2 \phi_r^*
\phi_r, \qquad r=1,2,
\end{equation}
%
\begin{equation}
\rho(x)=-\sum\limits_{r=1}^2 g_r \phi_r^* \phi_r.
\end{equation}
Poincar\'e-invariance of Yukawa model leads to existence of ten
conserved quantities which are a 4-momentum $P^\mu$ and 4-angular
momentum $M^{\lambda\sigma}$.

For the Lagrangian (\ref{L}) these expressions were found in
\cite{du}:
%
\begin{eqnarray}
\label{1.3}
P^\mu(t)=\sum\limits_{r=1}^2\int d^3x T^{0\mu}_r
(x)|_{x^0=t}-\eta^{0\mu}\int d^3x\int d^4 x'
\rho(x)K(x-x')\rho(x')|_{x^0=t}  \nonumber\\
{}-\frac{1}{2}\int\ dx^4
\int d^4 x' \Xi (x^0-t, x'^0-t)\rho(x)\{\partial^\nu
K(x-x')\}\rho(x'),
\end{eqnarray}
%
\begin{eqnarray}\label{1.4}
M^{\lambda\sigma}(t)=\sum\limits_{r=1}^2\int d^3 x
T_r^{0[\lambda}(x)x^{\sigma]}|_{x^0=t}-\int d^3x\int d^4 x' \rho(x)
\eta^{0[\lambda}x^{\sigma]}
K(x-x')\rho(x')|_{x^0=t} \nonumber\\
{}-\frac{1}{2}\int\ dx^4 \int d^4 x' \Xi (x^0-t, x'^0-t)\rho(x)\{
\partial^{[\lambda} K(x-x')x^{\sigma]}\}\rho(x').
\end{eqnarray}
Here $a^{[\mu}b^{\nu]}=a^\mu b^\nu-a^\nu b^\mu$ та $\Xi(t,s)\equiv
\Theta(t)\Theta(-s)-\Theta(-t)\Theta(s)=\frac{1}{2}(\mathrm{sign}
(t) - \mathrm{sign} (s))$, where $\Theta(t)$ -- is the Heaviside
step function, and
%
\begin{equation}
\label{1.2.1} T_r^{\mu \nu}=\{(\partial^\mu \phi_r^*)(\partial^\nu
\phi_r)+(\partial^\nu \phi_r^*)(\partial^\mu \phi_r)\}-\eta^{\mu\nu} L_r
\end{equation}
is the energy-momentum tensor for a free field $\phi_r (x)$.

For further calculation it is convenient to transform each complex
field into a pair of real fields: $\phi_{r\alpha}(x)$ ($r=1,2$;
$\alpha=1,2$):
%
\begin{equation}
\label{1.10}
\phi_r=\frac{1}{\sqrt{2}}(\phi_{r1}+i\phi_{r2}), \quad   \phi_r^*=\frac{1}{\sqrt{2}}(\phi_{r1}-i\phi_{r2}),
\end{equation}
and, for brevity, we replace the multi subscript $r\alpha$ by single
subscript $a$ ($a=\overline{1,4}$). Than we have
%
\begin{equation}
\label{1.10.1}
\rho(x)=-\frac{1}{2}\sum\limits_a g_a \phi_a^2(x).
\end{equation}

The transition to the Hamiltonian description was realized in
\cite{du} by means of the Hamiltonization procedure for nonlocal
Lagrangians \cite{los}. This transition is built as a perturbative
scheme with the usage of the momentum representation for fields that
in the first-order approximation (in a coupling constant squared)
has the simple form:
%
\begin{equation}
\label{1.11} \phi_a(x)=\frac{1}{(2\pi)^{3/2}}\sum\limits_{A=\pm}\int
\frac{d^3 k}{\sqrt{2k_{a0}}}\ a_a^A(\mathbf{k})e^{iAk_ax}, \quad
a=1,...,4,
\end{equation}
where $k=\{k_{a0},\mathbf{k}\}$, $k_{a0}=\sqrt{m_a^2+\mathbf{k}^2}$,
$\mathbf{k}=\{k^i, i=1,2,3\}$, and quantities $a_a^A$ are the
amplitudes of normal field modes which upon quantization become the
particles creation ($A=+$) and annihilation ($A=-$) operators.

For the generators of time translations $H=P^0$ (the Hamiltonian)
and space translations $\mathbf{P}=\{P^i,\ i=1,2,3\}$ (the momentum)
it were found the following expressions \cite{du}:
$$H=H_{free}+H_{int},\qquad
\mathbf{P}=\mathbf{P}_{free},$$
where
%
\begin{eqnarray}
\label{1.6.1}
H_{free}&=&\frac{1}{2}\sum\limits_a\sum\limits_A\int d^3k\,
k_{a0}a^A_a(\mathbf{k})a^{-A}_a(\mathbf{k}),\\
\label{1.7.1}
\mathbf{P}_{free}&=&\frac{1}{2}\sum\limits_a\sum\limits_A\int d^3k\,
\mathbf{k}a^A_a(\mathbf{k})a^{-A}_a(\mathbf{k}),
\end{eqnarray}
\begin{equation}\label{1.8}
H_{int}=\frac{1}{2}\sum\limits_{ab}\sum\limits_{ABCD}\int\!\!
d^3k\,d^3q\,d^3u\,d^3v \,
T^{ABCD}_{ab}(\mathbf{k},\mathbf{q},\mathbf{u},\mathbf{v})
a^A_a(\mathbf{k})a^B_a(\mathbf{q})a^C_b(\mathbf{u})a^D_b(\mathbf{v}),
\end{equation}
and where
%
\begin{equation}\label{13}
T^{ABCD}_{ab}(\mathbf{k},\mathbf{q},\mathbf{u},\mathbf{v})
=-\frac{g_a g_b}{16(2\pi)^3}\frac{\delta(A\mathbf{k}+B\mathbf{q}+C\mathbf{u}+D\mathbf{v})}
{\sqrt{k_{a0}q_{a0}u_{b0}v_{b0}}}\widetilde{K}(Ak_a+Bq_a),
\end{equation}
%
\begin{equation}
\widetilde{K}(k)=\int d^4 x\, e^{-ik\cdot x}K(x).
\end{equation}

In the expressions (\ref{1.6.1}) and (\ref{1.7.1}) a summation
over \emph{A} can be performed:
%
\begin{eqnarray}
\label{1.34}
H_{free}&=&\sum\limits_a\int d^3k\,
k_{a0}a^+_a(\mathbf{k})a^-_a(\mathbf{k}),\\
\label{1.33}
\mathbf{P}_{free}&=&\sum\limits_a\int d^3k\,
\mathbf{k}a^+_a(\mathbf{k})a^-_a(\mathbf{k}).
\end{eqnarray}
The translation generators \emph{H} and \textbf{P} must be
supplemented with  generators of the Lorentz group for further
Poincar\'e-invariance examination of the system. These generators form
into the 4-angular momentum.


\section {ANGULAR MOMENTUM}

\textit{Zero-order approximation}\\

We find space components of angular momentum in zero-order
approximation at first. The energy-momentum tensor for complex
scalar fields (\ref{1.2.1}) must be substituted into the first term
of (\ref{1.4}). Thus we obtain the expression (with $i,j=1,2,3$):
%
\begin{equation}
\label{1.9}
M^{ij}_{free}\equiv M^{ij}_{(0)}=\sum\limits_r\int d^3x \{\dot{\phi_r}^*(x^j \partial^i \phi_r -x^i \partial^j \phi_r)
+\dot{\phi_r}(x^j \partial^i \phi_r^* -x^i \partial^j \phi_r^*) \}.
\end{equation}
We proceed to real fields (\ref{1.10}), rename subscripts
($r\alpha\rightarrow a$), take into consideration the representation
(\ref{1.11}), and obtain the components of angular momentum vector:
%
\begin{equation}
M^k_{(0)}\equiv \frac{1}{2}\varepsilon^k_{\; ij}M^{ij}_{(0)}=
\frac{i}{2}\varepsilon^k_{\; ij}\sum\limits_a \sum\limits_A \int d^3
kAa_a^A(\mathbf{k})k^i \partial^j a_a^{-A}(\mathbf{k}),\qquad k=1, 2, 3.
\end{equation}
After summation over \emph{A} we have:
%
\begin{equation}\label{1.30}
M^k_{free}\equiv M^k_{(0)}=i\varepsilon^k_{\; ij}\sum\limits_a \int d^3 k\,
a_a^+(\mathbf{k})k^i \partial^j a_a^-(\mathbf{k});
\end{equation}
here $\partial^ia(\mathbf{k})=\partial a(\mathbf{k})/\partial k_i$ etc.\\

\noindent
\textit{Fist-order approximation}\\

Using (\ref{1.4}) and (\ref{1.2.1}) the first correction for
angular momentum can be written as follows:
%
\begin{equation}
M^{ij}_{(1)}=
\int d^4 x \int d^4 x'\,\Xi (x^0, x'^0)\rho(x') K(x-x')[\partial^i
\rho(x)x^j-\partial^j \rho(x)x^i].
\end{equation}
Then we transform each complex field into a pair of real fields
(\ref{1.10}), take into consideration eqs. (\ref{1.10.1}) and
(\ref{1.11}) and arrive at the formula:
%
\begin{multline}
M^{ij}_{(1)}=4i\sum\limits_{ab}\sum\limits_{ABCD}\int
d^3k\, d^3q\, d^3u\,d^3v\,
S^{ABCD}_{ab}(\mathbf{k},\mathbf{q},\mathbf{u},\mathbf{v})a^A_a(\mathbf{k})a^B_a(\mathbf{q})
\\
\ \quad\times \{u^ja^D_b(\mathbf{v})\partial^i a^C_b(\mathbf{u})
-u^ia^D_b(\mathbf{v})\partial^j a^C_b(\mathbf{u})
+v^ja^C_b(\mathbf{u})\partial^i a^D_b(\mathbf{v})
-v^ia^C_b(\mathbf{u})\partial^j a^D_b(\mathbf{v})\},
\end{multline}
where the kernel
%
\begin{multline}
S^{ABCD}_{ab}(\mathbf{k},\mathbf{q},\mathbf{u},\mathbf{v})
=\frac{g_a
g_b}{16(2\pi)^3}\frac{\delta(A\mathbf{k}+B\mathbf{q}+C\mathbf{u}+D\mathbf{v})}
{\sqrt{k_{a0}q_{a0}u_{b0}v_{b0}}}
\\
\times {\cal
P}\frac{\widetilde{K}(Ak_a+Bq_a)-\widetilde{K}(Cu_b+Dv_b)}{Ak_{a0}+Bq_{a0}+Cu_{b0}+Dv_{b0}}
\end{multline}
was found in \cite{du}\footnote{One of the authors (A.D.) asks
pardon for an error made in Eq. (5.22) of Ref. \cite{du} where the
mistaken factor 1/16 is to be read as 1/8.}.


\section {CENTRE-OF-MASS INTEGRAL}

\emph{Zero-order approximation}\\

Similarly to the angular momentum we find an expression for the integral
of centre-of-mass (it corresponds to pure Lorentz transformations) in
zero-order approximation. For this purpose the expression (\ref{1.2.1})
for the energy-momentum tensor must be substituted into the
first term of (\ref{1.4}) (where we assign $\lambda=0$, $\mu=i$ for superscripts):
%
\begin{equation}
K^i_{(0)}\equiv M^{0i}_{(0)}=\sum \limits_r \int d^3x\,
\{2\dot{\phi_r}^*\dot{\phi_r}\: x^i-x^0(\dot{\phi_r}^*\partial^i \phi_r+\partial^i \phi_r^*\dot{\phi_r})\}.
\end{equation}
After proceeding to real fields and some transformations and
substitutions we obtain:
%
\begin{equation}
K^i_{(0)}=\frac{i}{2}\sum \limits_a \sum \limits_A \int d^3k\,
A a^A_a(\mathbf{k})\{k_{a0}\partial^i a^{-A}_a(\mathbf{k})
+\frac{k^i}{k_{a0}}a^{-A}_a(\mathbf{k})\}.
\end{equation}
The second term contains two equal components with opposite
signs. Thus, after summation over \emph{A}, the only first term
survives:
%
\begin{equation}\label{1.31}
K^i_{free}\equiv K^i_{(0)}=\frac{i}{2}\sum \limits_a \int d^3k\,
k_{a0} a^+_a(\mathbf{k})\stackrel{\leftrightarrow}{\partial^i} a^-_a(\mathbf{k}),
\end{equation}
where $ a \stackrel{\leftrightarrow}{\partial^i} b\equiv
a\partial^i b-(\partial^i a)b
$.\\

\noindent
\emph{Fist-order approximation}\\

According to (\ref{1.4}),
the centre-of-mass integral in the fist-order approximation is:
$$
K^i_{(1)}=K^i_{int}+K^i_{nc},
$$
where
%
\begin{multline}\label{1.32}
K^i_{int}=-\frac{i}{2}\sum\limits_{ab}\sum\limits_{ABCD}A\int d^3k\,
d^3q\, d^3u\,d^3v\,
T^{CDAB}_{ba}(\mathbf{u},\mathbf{v},\mathbf{k},\mathbf{q})
\\
\times\{\partial^i a^A_a(\mathbf{k})
-\frac{k^i}{2k_{a0}^2}a^A_a(\mathbf{k})\}
a^B_a(\mathbf{q})a^C_b(\mathbf{u})a^D_b(\mathbf{v}),
\end{multline}
\begin{multline}
K^i_{nc}=4i\sum\limits_{ab}\sum\limits_{ABCD}\int d^3k\, d^3q\,
d^3u\,d^3v\,
S^{ABCD}_{ab}(\mathbf{k},\mathbf{q},\mathbf{u},\mathbf{v})a^A_a(\mathbf{k})a^B_a(\mathbf{q})
\\
\times \{u_{b0}a^D_b(\mathbf{v})\partial^i a^C_b(\mathbf{u})
+v_{b0}a^C_b(\mathbf{u})\partial^i a^D_b(\mathbf{v})
+\frac{a^D_b(\mathbf{v})a^C_b(\mathbf{u})}{2}\left(\frac{u^i}{u_{b0}}-\frac{v^i}{v_{b0}}\right)\}.
\end{multline}


\section {CHANGE OF VARIABLES. CANONICAL QUANTIZATION}

As it was shown in \cite{du}, the variables $a^+_a, a^-_a$ are
non-canonical. We make a transition to canonical variables
$\underline{a}$ that satisfy the Poisson bracket relations:
%
\begin{equation}
\{\underline{a}^-_a(\mathbf{k}),
\underline{a}^+_b(\mathbf{q})\}=i\delta_{ab}\delta(\mathbf{k}-\mathbf{q}).
\end{equation}
(other Poisson brackets are equal to zero). These variables are related to the original ones by
the approximated formula \cite{du}:
%
\begin{multline}
a^A_a(\mathbf{k})=\underline{a}^A_a(\mathbf{k})+\frac{A}{2}\sum\limits_b\sum\limits_{BCD}\int d^3q\, d^3u\,d^3v\,
S^{{-}ABCD}_{ab}(\mathbf{k},\mathbf{q},\mathbf{u},\mathbf{v})\\
\times\underline{a}^B_a(\mathbf{q})\underline{a}^C_b(\mathbf{u})\underline{a}^D_b(\mathbf{v})
+o(g^2),
\end{multline}
where the symbol $(g^2)$ denotes terms of higher order than $g_a^2$ and $g_ag_b$.
Thus the expressions for the angular momentum and for the centre-of-mass integral
in terms of new variables $\underline{a}$ can be written as follows:
%
\begin{eqnarray}\label{1.13}
&&\hspace{-2ex} \mathbf{M}=\mathbf{M}_{(0)}[a]+\mathbf{M}_{(1)}[a]
=\mathbf{M}_{free}[a]+\mathbf{M}_{nc}[a]=\mathbf{M}_{free}[\underline{a}]+o(g^2),\\
\label{1.14} &&\hspace{-2ex}
 \mathbf{K}=\mathbf{K}_{(0)}[a]+\mathbf{K}_{(1)}[a]
=\mathbf{K}_{free}[a]+\mathbf{K}_{int}[a]+\mathbf{K}_{nc}[a]
=\mathbf{K}_{free}[\underline{a}]+\mathbf{K}_{int}[\underline{a}]+o(g^2).
\end{eqnarray}
Henceforth, for convenience, we do not underscore new variables,
i.e. we rename $\underline{a}\rightarrow a$. Then the final
expressions for the angular momentum and for the centre-of-mass
integral in the momentum representation can be represented by the
formulae (\ref{1.30}), (\ref{1.31}) and (\ref{1.32}). Together with
(\ref{1.33}), (\ref{1.34}) and (\ref{1.8}) this yields a dynamical
basis for the system of two interacting scalar fields under
consideration.

Let us perform the canonical quantization. Then the variables
$a^+_a$ are the creations operators and $a^-_a$ are the annihilation
operators. The normal ordering of products of these operators is
understood. Poisson brackets should be replaced by quantum
commutators:
%
\begin{equation}
\{A, B\}\longrightarrow-i[A, B].
\end{equation}
For the operators $a^\pm$ standard commutational relations hold:
%
\begin{equation}
\label{kom}
[a^+_a(\mathbf{k}),a^+_b(\mathbf{q})]=[a^-_a(\mathbf{k}),a^-_b(\mathbf{q})]=0,\quad
[a^-_a(\mathbf{k}),a^+_b(\mathbf{q})]=\delta_{ab}\delta(\mathbf{k}-\mathbf{q}).
\end{equation}
%


\section {POINCAR\'E-INVARIANCE}

It is necessary to make sure that approximated expressions for operators
$H$, $\mathbf{P}$,  $\mathbf{M}$ and $\mathbf{K}$ do satisfy
the Poincar\'e algebra relations to get to know that the system possesses
a Poincar\'e-invariance. Scilicet the following expressions must be
verified \cite{haida}:
%
\begin{eqnarray}
&[P^i, H]=[P^i, H_{free}]+[P^i, H_{int}]=0,\nonumber\\
&[ M^i, H ] = [ M^i, H_{free}] + [ M^i, H_{int}]=0, \nonumber \\
&[ P^i, P^j ]=0,\quad [M^i,P^j]=i\varepsilon^{ij}_{~~k} P^k,\quad [M^i,M^j]=i\varepsilon^{ij}_{~~k} M^k,
\nonumber\\
\label{alg}
&[M^i,K^j]=[M^i,K^j_{free}]+[M^i,K^j_{int}]=i\varepsilon^{ij}_{~~k} K^k,
\nonumber\\
&[K^i,H]=[K^i_{free},H_{free}]+[K^i_{free},H_{int}]+[K^i_{int},H_{free}]+o(g^2)\simeq iP^i,
\nonumber\\
&[K^i,P^j]=[K^i_{free},P^j]+[K^i_{int},P^j]=i\delta^{ij}H,
\nonumber\\
&[K^i,K^j]=[K^i_{free},K^j_{free}]+[K^i_{free},K^j_{int}]+[K^i_{int},K^j_{free}]
+o(g^2)\simeq=-i\varepsilon^{ij}_{~~k} M^k.
\end{eqnarray}

It is easy to verify that commutation relations for free-field terms
are valid. Let us present a calculation of commutation relations in
the firs-order approximation in the coupling constant squared.

\subsection {Calculation of $[\textbf{P}, H_{int}]$ and $[M, H_{int}]$}

Let us mention two remarks that concern to calculation
of all commutators with interaction terms. The first concerns
to normal ordering of products of the creation and annihilation
operators. It is easy to verify that the normal ordering is preserved
at the every step of calculation below, regardless of the case
how operators are ordered -- explicitly or not.
Therefore for simplicity of the description we will to consider
interaction terms of generators ordered implicitly. The second remark
concern to superficial terms that arise during the calculation.
Those terms will be omitted so far as they give a zero action in
the Fock space.

Let us show the calculation of the first commutator in details
here:
\begin{multline}
[\mathbf{P}, H_{int}]=\frac{1}{2}\sum\limits_{abc}
\sum\limits_{ABCD}\int d^3p \,d^3k \, d^3q \, d^3u \, d^3v \,
T_{ab}^{ABCD}(\mathbf{k}, \mathbf{q}, \mathbf{u}, \mathbf{v})
\\
\times\mathbf{p} [a^+_c(\mathbf{p})a^-_c(\mathbf{p}),
a^A_a(\mathbf{k})a^B_a(\mathbf{q})a^C_b(\mathbf{u})a^D_b(\mathbf{v})]
\nonumber
\end{multline}
For brevity it is convenient to unify the integration variables
$\textbf{p}$, $\textbf{k}$ \dots and subscripts \emph{a},
\emph{b}, \emph{c} in the common subscripts \emph{p}, \emph{k},
\dots: $a^+_c(\textbf{p})\equiv a^+_p$, \dots . Let us consider
a commutator in r.-h.s. of equality quoted above and execute
some simplification in it. By consecutive permutations the
operator product $a^+ a^-$ with operators $a^A_k\dots a^D_v$
one obtains:
\begin{eqnarray}
[a^+_p a^-_p, a^A_k a^B_q a^C_u a^D_v]=&
-a^A_k a^B_q a^C_u a^D_p \delta_{pv}D
-a^A_ka^B_q a^C_p a^D_v\delta_{pu}C
\nonumber\\
&-a^A_k a^B_pa^C_u a^D_v\delta_{pq}B
-a^A_p a^B_q a^C_ua^D_v\delta_{pk}A.
\nonumber
\end{eqnarray}
Here we use the short notation:
$\delta_{pk}\equiv\delta_{ca}\delta(\mathbf{p}-\mathbf{k})$
etc.


Integrating this expression out and using some properties of
$\delta$-function we receive:
\begin{multline}
\int d^3p \, d^3k \, d^3q \, d^3u \, d^3v \,\mathbf{p}\,[a^+_p a^-_p, a^A_k
a^B_q a^C_u a^D_v]\\
=\int d^3k \, d^3q \, d^3u \, d^3v
\,(A\mathbf{k}+B\mathbf{q}+C\mathbf{u}+D\mathbf{v})a^A_k a^B_q
a^C_u a^D_v.\nonumber
\end{multline}
It is worth mentioning that the expression (\ref{13}) for $T^{ABCD}_{ab}$
contains a $\delta$-function. Thus we have
$$
\int d^3k \, d^3q \, d^3u \, d^3v \, (A\mathbf{k}+B\mathbf{q}+C\mathbf{u}+D\mathbf{v})
\delta(A\mathbf{k}+B\mathbf{q}+C\mathbf{u}+D\mathbf{v})\dots=0,
$$
where the following property of $\delta$-function is taken into
consideration: $\int dx\, \delta(x)xf(x) = 0$ for arbitrary
function $f(x)$ that is regular in $x=0$. Thus:
$$
[\mathbf{P}, H_{int}]=0.
$$

The second commutator can be calculated by analogy. The terms
that contain the derivative of $\delta$-function will occur
there. In this case the differential operation should be
displaced onto a one of operators $a^\pm$ (omitting the
superficial terms). Finally we obtain that the second
commutator is equal to zero:
$$
[\mathbf{M}, H_{int}]=0.
$$

\subsection {Calculation of $[H_{free}, \mathbf{K}_{int}]$ and $[H_{int}, \mathbf{K}_{free}]$}

Since free-field generators satisfy the Poincar\'e algebra
(\ref{alg}) it is necessary to prove the equality:
%
\begin{equation}
\label{6.1} [H_{free},
\mathbf{K}_{int}]-[\mathbf{K}_{free}, H_{int}]=0.
\end{equation}
In order to simplify calculations we proceed from operators
$a^\pm$ to operators $b^\pm$ by means of the relation:
%
\begin{equation}
\label{b}
a_k^\pm=\sqrt{k_0}b_k^\pm.
\end{equation}
In the new notation:
%
\begin{equation}
 H_{free}=\int d^3k\, k_0^2 b^+_k b^-_k,
\end{equation}
%
\begin{equation}
\label{kf}
\mathbf{K}_{free}=\frac{i}{2}\int d^3k\, k_0^2 b^+_k
\stackrel{\leftrightarrow}{\nabla} b^-_k,
\end{equation}
%
\begin{equation}
\label{6.1.1}
H_{int}=\sum\limits_{ABCD}\int d^3k\, d^3q\, d^3u\,d^3v\,
\Pi^{ABCD}_{kquv}b^A_k b^B_q b^C_u b^D_v,
\end{equation}
%
\begin{equation}
\label{6.1.2}
\mathbf{K}_{int}=-i\!\!\sum\limits_{ABCD}\int d^3k\, d^3q\,
d^3u\,d^3v\,
D\Pi^{ABCD}_{kquv}b^A_k b^B_q b^C_u \nabla b^D_v,
\end{equation}
(integrating by $d^3k \, d^3q \, d^3u \, d^3v$ include the
summation over subscripts $a, b, c, d$ too; this subscripts are
not showed explicitly in the formulae). Here the kernel
$\Pi^{ABCD}_{kquv}$ is similar by structure to the defined above
kernel $T^{ABCD}_{ab}(\mathbf{k}, \mathbf{q}, \mathbf{u},
\mathbf{v})$ (\ref{13}):
%
\begin{eqnarray}\label{Kern}
\Pi^{ABCD}_{kquv}
&\equiv&\Pi^{ABCD}_{ab}(\mathbf{k}, \mathbf{q}, \mathbf{u}, \mathbf{v})
=\frac{1}{2}\sqrt{k_{a0} q_{a0} u_{b0} v_{b0}}\,
T^{ABCD}_{ab}(\mathbf{k}, \mathbf{q}, \mathbf{u}, \mathbf{v})\nonumber\\
&=&-\frac{g_a g_b}{32(2 \pi)^3} \delta (A\mathbf{k}+B\mathbf{q}+C\mathbf{u}+D\mathbf{v})\widetilde{K}(Ak_a+Bq_a).
\end{eqnarray}
Similarly to $T^{ABCD}_{ab}(\mathbf{k}, \mathbf{q}, \mathbf{u},
\mathbf{v})$ this kernel possesses symmetry properties:
%
\begin{eqnarray}
\Pi^{BACD}_{ab}(\mathbf{q}, \mathbf{k}, \mathbf{u},
\mathbf{v})&=&\Pi^{ABCD}_{ab}(\mathbf{k}, \mathbf{q},
\mathbf{u}, \mathbf{v}),\nonumber\\
\Pi^{ABCD}_{ab}(-\mathbf{k}, -\mathbf{q}, -\mathbf{u},
-\mathbf{v})&=&\Pi^{-A-B-C-D}_{ab}(\mathbf{k}, \mathbf{q},
\mathbf{u}, \mathbf{v})=\Pi^{ABCD}_{ab}(\mathbf{k}, \mathbf{q},
\mathbf{u}, \mathbf{v}),\label{11.1}\\
\Pi^{-A-BCD}_{ab}(\mathbf{k}, \mathbf{q}, \mathbf{u},
\mathbf{v})&=&\Pi^{ABCD}_{ab}(\mathbf{k}, \mathbf{q},
-\mathbf{u}, -\mathbf{v}),\nonumber\\
\Pi^{AB-CD}_{ab}(\mathbf{k}, \mathbf{q}, -\mathbf{u},
\mathbf{v})&=&\Pi^{ABC-D}_{ab}(\mathbf{k}, \mathbf{q}, \mathbf{u},
-\mathbf{v})=\Pi^{ABCD}_{ab}(\mathbf{k}, \mathbf{q}, \mathbf{u},
\mathbf{v})\label{11.1a}
\end{eqnarray}
which are important for calculations. Summing over \emph{A, B,
C, D} in (\ref{6.1.1}), (\ref{6.1.2}) up and using some
properties (\ref{11.1}) of $\Pi ^{ABCD}_{kquv}$ yields:
%
\begin{eqnarray}
\label{6.1.3} H_{int}=\int d^3k...d^3v&\{&\Pi^{----}_{kquv}b^-_k
b^-_q b^-_u b^-_v
\nonumber\\
&&+2(\Pi^{+---}_{kquv}+\Pi^{--+-}_{uvkq})b^+_k b^-_q b^-_u b^-_v
\nonumber\\
&&+(\Pi^{++--}_{kquv}+\Pi^{--++}_{uvkq}+4 \Pi^{+-+-}_{kuqv})b^+_k
b^+_q b^-_u b^-_v\nonumber\\
&&+2(\Pi^{+++-}_{kquv}+\Pi^{+-++}_{uvkq})b^+_k b^+_q b^+_u b^-_v
\nonumber\\
&&+\Pi^{++++}_{kquv}b^+_k b^+_q b^+_u b^+_v\},
\end{eqnarray}
%
\begin{eqnarray}
\label{6.1.4} \mathbf{K}_{int}&=&i\int
d^3k...d^3v\,\{\Pi^{----}_{kquv} b^-_k b^-_q b^-_u \nabla_v b^-_v
\nonumber\\
&&{}+2\Pi^{+---}_{kquv} b^+_k b^-_q b^-_u \nabla_v b^-_v+
\Pi^{--+-}_{uvkq}b^+_k \nabla_q b^-_q b^-_u b^-_v -
\Pi^{--+-}_{uvkq}\nabla_k b^+_k b^-_q b^-_u b^-_v
\nonumber\\
&&{}+ (\Pi^{++--}_{kquv}+ 2 \Pi^{+-+-}_{kuqv})b^+_k b^+_q b^-_u
\nabla_v b^-_v - 2i \Pi^{+-+-}_{kuqv}b^+_k \nabla_q b^+_q b^-_u
b^-_v - i \Pi^{--++}_{uvkq}\nabla_k b^+_k b^+_q b^-_u b^-_v
\nonumber\\
&&{}+ \Pi^{+++-}_{kquv}b^+_k b^+_q b^+_u \nabla_v b^-_v -
\Pi^{+++-}_{kquv}b^+_k b^+_q \nabla_u  b^+_u b^-_v -2
\Pi^{+-++}_{uvkq}\nabla_k b^+_k b^+_q b^+_u b^-_v
\nonumber\\
&&{}- \Pi^{++++}_{kquv} b^+_k b^+_q b^+_u \nabla_v b^+_v\}.
\end{eqnarray}

 Thereafter we calculate commutators for items with fixed
number of creation and annihilation operators separately. Let
us calculate one of them.

It is worth to mention at first the equalities:
%
\begin{equation}
\label{bb}
[b^-_k, b^+_q]=\frac{\delta_{kq}}{q_0}, \, \:
[\nabla_k b^-_k,
b^+_q]=\frac{ \nabla_k \delta_{kq}}{q_0},
\, \: [b^-_k,
\nabla_q b^+_q]=\frac{\nabla_q \delta_{kq}}{k_0}.
\end{equation}

Let us find commutator $H_{free}$ with the first line of
expression for $\mathbf{K}_{int}$ (\ref{6.1.4}):
%
\begin{multline}\label{47}
\int d^3p\,d^3k\dots d^3v\,[p^2_0 b^+_p b^-_p, i \Pi
^{----}_{kquv}b^-_k b^-_q b^-_u \nabla_v b^-_v]\\
=i\int d^3k\dots d^3v\,(k_0+q_0+u_0+v_0)\nabla_v \Pi
^{----}_{kquv}b^-_k b^-_q b^-_u b^-_v,
\end{multline}
and a commutator $\mathbf{K}_{free}$ with the first line of
expression for $H_{int}$ (\ref{6.1.3}):
%
\begin{eqnarray}\label{48}
\frac{i}{2}\int d^3p\,d^3k\dots d^3v\,[p^2_0(b^+_p \nabla_p b^-_p- \nabla_p b^+_p b^-_p),
\Pi^{----}_{kquv}b^-_k b^-_q b^-_u b^-_v]
\nonumber\\
=i\int d^3k\dots d^3v\,(k_0\nabla_k \Pi^{----}_{kquv}+q_0\nabla_q \Pi
^{----}_{kquv}
\nonumber\\
{}+u_0\nabla_u \Pi ^{----}_{kquv}+v_0\nabla_v \Pi
^{----}_{kquv})b^-_k b^-_q b^-_u b^-_v.
\end{eqnarray}
Now it is necessary to show that the difference of the
integrals (\ref{47}) and (\ref{48}) is equal to zero, i.e.:
%
\begin{multline}\label{49}
(k_0+q_0+u_0+v_0)\nabla_v \Pi^{----}_{kquv}
\\
{}-(k_0\nabla_k\Pi^{----}_{kquv}+q_0\nabla_q \Pi^{----}_{kquv}
{}+u_0\nabla_u \Pi^{----}_{kquv}+v_0\nabla_v \Pi^{----}_{kquv} )=0.
\end{multline}
To show this let us take into consideration the structure of
kernel (\ref{Kern}), namely
%
\begin{multline}\label{Struct}
\Pi^{ABCD}_{kquv}\propto\delta(A\mathbf{k}+B\mathbf{q}+C\mathbf{u}+D\mathbf{v})
\widetilde{K}(Ak+Bq),\\
\mbox{where}~~~~ \widetilde{K}(Ak+Bq)\equiv\widetilde{K}[(Ak+Bq)^2].
\end{multline}
It is easy to approve the identity:
%
\begin{equation}
\label{K}
k_0 \nabla_k \widetilde{K}(k\pm q)+q_0 \nabla_q \widetilde{K}(k\pm q)=0.
\end{equation}
Expanding the expression in l.-h.s. of eq. (\ref{49})
yields:
\begin{multline}
(k_0+q_0+u_0+v_0)\nabla\delta(\mathbf{k+q+u+v})\widetilde{K}(k+q)\\
-k_0\nabla\delta(\mathbf{k+q+u+v})\widetilde{K}(k+q)-
k_0\delta(\mathbf{k+q+u+v})\nabla_k \widetilde{K}(k+q)\\
-q_0\nabla\delta(\mathbf{k+q+u+v})\widetilde{K}(k+q)-
q_0\delta(\mathbf{k+q+u+v}) \nabla_q \widetilde{K}(k+q)\\
-u_0\nabla\delta(\mathbf{k+q+u+v})\widetilde{K}(k+q)-
v_0\nabla\delta(\mathbf{k+q+u+v})\\
=-\delta(\mathbf{k+q+u+v})(k_0 \nabla_k \widetilde{K}(k+q)+ q_0
\nabla_q \widetilde{K}(k+q))=0, \nonumber
\end{multline}
thus the equality (\ref{49}) is true.

The commutators of terms with other rates of creation
and annihilation operators can be calculated by analogy.
Thus we have accomplished a proof of the equality (\ref{6.1}).

\subsection {Calculation of $[P^i, K^j_{int}]$ and $[M^k, K^j_{int}]$}

Upon calculating these commutators it is convenient to use
simplification proposed in the previous paragraph. Then
expressions for components of the momentum and the  angular momentum are:
$$
P^i=\int d^3k\, p^ip_0 b^+_p b^-_p
$$
and:
$$
M^k=i\varepsilon^k_{~ij}\sum \limits_a \int d^3p \, p^i p_0 b^+_p \partial^j b^-_p.
$$
After simple calculations we obtain:
$$
[P^i, K^j_{int}]=i \delta^{ij} H_{int}.
$$
As far as the second commutator is concerned,
it is necessary to prove the equality:
%
\begin{equation}
\label{mk}
[M^k, K^l_{int}]=i\varepsilon^{kl}_{~m}K^m_{int}.
\end{equation}
Let us show that this equality is fulfilled separately
for items with fixed numbers of creation and annihilation
operators in the expression $K_{int}^j$ (\ref{6.1.4}). For
example, for items with annihilation operators only we have:
\begin{multline}
\varepsilon^k_{~ij}\int d^3p\,p^i p_0
\Pi^{----}_{kqu\partial^lv}[b^+_p\partial^j b^-_p,b^-_k b^-_q b^-_u b^-_v]\\
=-\varepsilon^k_{ij} \partial^l_v \Pi^{----}_{kquv}(k^i\partial^j b^-_k b^-_q b^-_u b^-_v
+q^i b^-_k \partial^j b^-_q b^-_u b^-_v + u^i b^-_k b^-_q \partial^j b^-_u b^-_v
+v^i b^-_k b^-_q b^-_u \partial^j b^-_v).
\end{multline}
Let us switch over the derivatives from operators onto kernel and
throw away the unimportant superficial terms. Owing to
antisymmetry of the factor $\varepsilon^k_{~ij}$ we receive the
simple expression:
%
\begin{equation}\label{ex}
\widehat{L}_k
\partial_v^l\Pi^{----}_{kquv}b^-_k b^-_q b^-_u b^-_v,
\end{equation}
where
$\widehat{L}^k\equiv
\varepsilon^k_{~ij}(k^i\partial_k^j+q^i\partial_q^j+u^i\partial_u^j+v^i\partial_v^j)$
is the infinitesimal rotational operator. Let us write down:
$$
\widehat{L}^k \partial_v^l\Pi^{----}_{kquv}
=\partial_v^l\widehat{L}_k\Pi^{----}_{kquv}
b^-_v+\left[\widehat{L}_k,\partial_v^l\right]\Pi^{----}_{kquv}.
$$
Since we have the rotary-invariant kernel the first term in r.-h.s.
is equal to zero. After calculating the commutator
$\left[\widehat{L}^k,\partial_v^l\right]=-\varepsilon^k_{~lj}\partial_v^j$,
we present the sought expression (\ref{ex}) as:
$$
-\varepsilon^k_{ij} \partial^j_v \Pi^{----}_{kquv}b^-_k b^-_q b^-_u b^-_v.
$$
It is easy to see that it gives whole contribution in the
r.-h.s. of (\ref{mk}) which contains annihilation
operators only.

Commutators of the momentum with other lines of the expression
(\ref{6.1.4}) for
$K_{int}^j$ (that contain the creation operators too) can be
found by analogy. Finally, we complete the proof of equality
(\ref{mk}).

\subsection {Calculation of $[K^i_{free},K^j_{int}]$ and $[K^i_{int},K^j_{free}]$}

Let us find $[K^i_{free},K^j_{int}]+[K^i_{int},K^j_{free}]$.

By analogy to previous paragraph, we use the expressions
(\ref{kf}) and (\ref{6.1.4}) for
$K^i_{free}$ and $K^j_{int}$ in this sum in terms of $b^\pm$
operators. Then, using the commutational relations
(\ref{bb}), both
commutators are calculated line-by-line.

For the first line of  (\ref{6.1.4}) the sum of commutators under consideration yields:
\begin{multline}
\int d^3k\dots d^3v \, \left\{(k_0 \partial^i_k \partial^j_v+ q_0
\partial^i_q \partial^j_v+u_0 \partial^i_u \partial^j_v
+v_0 \partial^i_v \partial^j_v\right.\\
\left.-k_0 \partial^j_k \partial^i_v -q_0 \partial^j_q \partial^i_v
+u_0 \partial^j_u \partial^i_v -v_0 \partial^j_v \partial^i_v)
\Pi^{----}_{kquv}\right\}b^-_k b^-_q b^-_u b^-_v. \nonumber
\end{multline}
Taking into consideration the structure of kernel (\ref{Struct})
and obvious equalities:
$$\partial^i_k \partial^j_v
\delta(\mathbf{k+q+u+v})\equiv\partial^i \partial^j
\delta(\mathbf{k+q+u+v})=
\partial^j_k \partial^i_v\delta(\mathbf{k+q+u+v})$$
etc., we arrive at the expression:
\begin{multline}
\int d^3k \, d^3q \, d^3u \, d^3v \, (k_0 \partial^j_v \delta(\mathbf{k+q+u+v}) \partial^i_k \tilde K(k+q)\\
+q_0 \partial^j_v \delta(\mathbf{k+q+u+v}) \partial^i_q \tilde
K(k+q)
-k_0 \partial^i_v \delta(\mathbf{k+q+u+v})\partial^j_k \tilde K(k+q)\\
-q_0 \partial^i_v \delta(\mathbf{k+q+u+v}) \partial^j_q\tilde
K(k+q)b^-_k b^-_q b^-_u b^-_v, \nonumber
\end{multline}
that is equal to zero owing to (\ref{K}). The other commutators of
terms with definite rates of creation and annihilation operators can
be calculated by analogy. Thus one receives:
$$
[K^i_{free},K^j_{int}]+[K^i_{int},K^j_{free}]=0.
$$

\section {PROBLEM OF  UNITARITY OF THE SCATTERING MATRIX}
\subsection {$S$-matrix construction}

For the construction of $S$-matrix we use the standard algorithm of
quantum mechanics \cite{Qm}.

All the above mentioned quantities are given in Schrodinger
representation, in which the field operators $a^\pm
_a(\mathbf{k})\equiv a^\pm _a(\mathbf{k},t=0)$. Let us go over
to the interaction representation. For this we write Heisenberg
equation (relatively free-field Hamiltonian) for creating and
annihilating operators:
$$
i\dot{a}^A_a(\textbf{k},t)=[a^A_a(\textbf{k},t), H_{free}].
$$
We solve it and receive:
$$
a^A_a(\textbf{k},t)\equiv
e^{iH_{free}t}a^A_a(\textbf{k})e^{-iH_{free}t}=e^{-iAk_{a0}t}
a^A_a(\textbf{k}).
$$
Using this expressions for calculating of interaction
Hamiltonian, we receive:
\begin{multline*}
H_{int}(t) \equiv e^{iH_{free}t}H_{int}e^{-iH_{free}t}\\
=\frac{1}{2}\sum\limits_{ab}\sum\limits_{ABCD}\int\!\!
d^3k\,d^3q\,d^3u\,d^3v \,
T^{ABCD}_{ab}(\mathbf{k},\mathbf{q},\mathbf{u},\mathbf{v})\\
\times e^{-i(Ak_{a0}+Bq_{a0}+Cu_{b0}+Dv_{b0})t} :a^A_a(\textbf{k})
a^B_a(\textbf{q}) a^C_b(\textbf{u}) a^D_b(\textbf{v}):.
\end{multline*}

Let us consider an adiabatic scattering matrix in the first-order
approximation:
\begin{eqnarray*}
S&=\lim \limits_{\alpha\rightarrow 0} S_\alpha(\infty,
-\infty)\equiv \lim \limits_{\alpha\rightarrow
0}\mathrm{T}e^{-i\int \limits_{-\infty}^\infty\!\!
dt'e^{-\alpha|t|}\! H_{int}(t')} \\
& \simeq 1-i\lim \limits_{\alpha\rightarrow 0}\int
\limits_{-\infty} ^\infty dt' e^{-\alpha|t|}H_{int}(t');
\end{eqnarray*}
here T denotes a chronological ordering. Note that the operator of
chronological ordering T does not play any role in the present
approximation. Thus the calculation of $S$ is straitforward and the
final expression for scattering matrix can be received:
\begin{multline}\label{11.10}
 S=1+\frac{i}{2}\sum \limits_{ab}\!\!\sum\limits_{ABCD} \int\!\!
d^3k\,d^3q\,d^3u\,d^3v \, \frac{g_a g_b}{16(2\pi)^3}
\frac{\delta^{(4)}(Ak_a{+}Bq_a{+}Cu_b{+}Dv_b)}
{\sqrt{k_{a0}q_{a0}u_{b0}v_{b0}}} \\
\times\widetilde{K}(Ak_a+Bq_a)
:a^A_a(\mathbf{k})a^B_a(\mathbf{q})a^C_b(\mathbf{u})a^D_b(\mathbf{v}):.
\end{multline}

\subsection {Proof of the unitarity}

Now it should be shown that $S$ (\ref{11.10}) is unitary operator in
current approximation. It means:
$$
SS^+\approx1+o(g^2).
$$
Scatterin matrix operator can be written as:
\begin{equation}
\label{11.4} S=I+iF,
\end{equation}
where $F$ up to $i$ factor is equal to the transition operator. It
matches in current approximation to a phase operator (following
Blokhintsev; \cite{Bl_p}). Accordingly with (\ref{11.4}), $S$ is a
unitary operator if $F$ is hermitian: $F=F^+$.

Let us show it is really so. For convenience we go over to
the operators \emph{b} introduced in Subsection 6.2. Then
\begin{equation}
\label{11.11} F=\frac{1}{2}\sum \limits_{ab} \sum \limits_{ABCD}
\int\!\! d^3k\,d^3q\,d^3u\,d^3v \,
Q^{ABCD}_{kquv}:b^A_{ak}b^B_{aq}b^C_{bu}b^D_{bv}:\, ,
\end{equation}
where $Q^{ABCD}_{kquv}=\delta
(Ak_{a0}+Bq_{a0}+Cu_{b0}+Dv_{b0})\Pi^{ABCD}_{kquv}$ is
expressed here in terms of the kernel
$\Pi^{ABCD}_{kquv}$ (\ref{Kern}) introduced in Subsection 6.2
and possessing the properties (\ref{11.1}) and (\ref{11.1a}).
In contrast to $\Pi^{ABCD}_{kquv}$, the kernel Q satisfies properties
(\ref{11.1}) only.

After summation over A, B, C, D in (\ref{11.11}) we receive:
\begin{eqnarray}
\label{11.9}
F&=&\sum \limits_{ab} \int d^3k... d^3v\,:\{Q^{----}_{kquv}b^-_{ak} b^-_{aq} b^-_{bu} b^-_{bv} \nonumber \\
&&+2Q^{+---}_{kquv}b^+_{ak} b^-_{aq} b^-_{bu} b^-_{bv}
+2Q^{--+-}_{uvkq}b^+_{bk} b^-_{bq} b^-_{au} b^-_{av} \nonumber \\
&&+Q^{++--}_{kquv}b^+_{ak} b^+_{aq} b^-_{bu} b^-_{bv}+
Q^{--++}_{uvkq}b^+_{bk} b^+_{bq} b^-_{au} b^-_{av}+
4 Q^{+-+-}_{kuqv}b^+_{ak} b^+_{bq} b^-_{au} b^-_{bv} \nonumber \\
&&+2Q^{+++-}_{kquv}b^+_{ak} b^+_{aq} b^+_{bu} b^-_{bv}+
2Q^{+-++}_{uvkq}b^+_{bk} b^+_{bq} b^+_{au} b^-_{av} \nonumber \\
&&+Q^{++++}_{kquv}b^+_{ak} b^+_{aq} b^+_{bu} b^+_{bv}\}:.
\end{eqnarray}

Hermiticity can be proved separately for some groups of terms in
this expression: for the first and the last lines, for the second
and the fourth, separately for two first terms in the third line,
and separately it can be shown a hermiticity for the third term in
this line.

For example we consider the sum of the first and the last lines in
expression (\ref{11.9}). Let us conjugate this sum:
\begin{multline}
\label{1.111}
(Q^{----}_{kquv}:b^-_{ak} b^-_{aq} b^-_{bu}
b^-_{bv}:+Q^{++++}_{kquv}:b^+_{ak} b^+_{aq} b^+_{bu}
b^+_{bv}:)^+ \\
=Q^{----}_{kquv}:b^+_{bv} b^+_{bu} b^+_{aq}
b^+_{ak}:+Q^{++++}_{kquv}:b^-_{bv} b^-_{bu} b^-_{aq} b^-_{ak}:
\end{multline}

In kernels Q we inverse all the signs A, B, C, D accordingly to
(\ref{11.1}). Using normal ordering we order the operators \emph{b}
by indices \emph{k, q, u, v} and receive the expression:
$$
Q^{++++}_{kquv}:b^+_{ak} b^+_{aq} b^+_{bu}
b^+_{bv}:+Q^{----}_{kquv}:b^-_{ak} b^-_{aq} b^-_{bu} b^-_{bv}:. \,
$$
It coincides with an outgoing expression (in brackets in the
left-hand side of (\ref{1.111})). Therefore the sum of the first and
the last lines in (\ref{11.9}) has the property of hermiticity.

By analogy one can prove the hermiticity of other sums of lines in
expression (\ref{11.9}). At last we receive that $F$ is hermitian
operator and it means that the  scattering matrix $S$ is unitary.

\section {CONCLUSIONS}

We have considered the scalar Yukawa-like model within the framework
of partially reduced field theory. The Lagrangian of the model is a
time-nonlocal functional, and a transition to the Hamiltonian
formalism is nontrivial problem. In the preceding work \cite{du} the
perturbative Hamiltonization procedure and a quantization of model
were proposed. This procedure leads to a loss of manifest covariance
of the description. The question whether the procedure preserves a
relativistic invariance of the model remained open so far.

One of the aims of our work is a proof of the Poincar\'e-invariance
of quantum Hamiltonian description of the reduced Yukawa-like model.
For this purpose, the Hamiltonian counterparts to nonlocal Noether
integrals of angular momentum and centre-of-mass of the system found
earlier \cite{du} have been built in the first-order approximation
in a coupling constant squared $g^2$ and then have been quantized
canonically. In order to a Poincar\'e-invariance be guaranteed these
operators together with the Hamiltonian and momentum of the system
obtained in \cite{du} must satisfy the commutation relations of
Poincar\'e algebra, at least with precision up to $g^2$. This is
indeed demonstrated in the paper.

Upon calculations of commutators it has been noticed that an
availability or absence of a normal ordering in the expressions for
generators does not influence the result of calculation of
commutators of this generators. Moreover, the commutational relation
remain preserved in the case when only separate items with fixed
number of creation and annihilation operators are retained in the
interaction terms of canonical generators, (i.e., separate lines
only in expressions (\ref{6.1.3}) and (\ref{6.1.4})). It can be
interpreted as if the Poincar\'e-invariance is held approximately on
the every finite sector of the Fock space of the model. It is
expected also that the Poincar\'e-invariance is preserved in
higher-order approximations in coupling constant though this is
difficult to realized even in the second-order (i.e., in $g^4$)
approximation.

The other aim of our work concerns to scattering matrix for
investigated model. The explicit construction of S-matrix is
realized and its approximate unitarity is proved. These results
allow one to extend the field of application of this model to a
scattering problem.

It is not surprise that problem with unitarity do not arise in the
present approach. The first reason is that we consider only
lower-order approximations. The second reason is that the present
model is close to the local Yukawa model rather than to typical
models in the nonlocal field theory.

Investigation of more realistic systems such as partially
reduced spinor electrodynamic is addressed to subsequent works.

\end{document}